\documentstyle[twoside,fleqn,espcrc2]{article}
\newcommand{\AmS}{{\protect\the\textfont2
  A\kern-.1667em\lower.5ex\hbox{M}\kern-.125emS}}

\title{Four-quark energies in SU(2) lattice Monte Carlo using a tetrahedral
geometry\thanks{This work is part of the EC Programme
``Human Capital and Mobility'' -- project number ERB-CHRX-CT92-0051
-- and was presented by A.M.Green.} }

\author{A.M.Green$^{\rm a}$, J. Lukkarinen$^{\rm a}$, P.
Pennanen\address{Research Institute for Theoretical Physics,
University of Helsinki, Finland \\
E-mail address: green@phcu.helsinki.fi}%
       , C.Michael\address{DAMTP, University of Liverpool, UK \\
E-mail address: cmi@liverpool.ac.uk}
   and
S.Furui\address{School of Science and Engineering, Teikyo University, Japan\\
E-mail address: furui@dream.ics.teikyo-u.ac.jp}}

\begin{document}

\begin{abstract}
 This contribution -- a continuation of the work in refs.\cite{GMP}-\cite{GMS}
 -- reports
on recent developments in the calculation and understanding of
4-quark energies generated using lattice Monte Carlo techniques.
\end{abstract}

\maketitle

\section{THE CALCULATION OF 4-QUARK ENERGIES IN A TETRAHEDRAL GEOMETRY USING
MONTE CARLO SIMULATION}

In refs.\cite{GMP2,GMS} -- the binding energies of four quarks in various
geometries (e.g. at the
corners of a rectangle or on a straight line) were calculated in
static-quenched-SU(2) on a lattice $16^3\times 32$ with $\beta=2.4$.
This showed that the 4-quark
binding energies  were greatest when two of the three possible partitions of
the
four quarks -- into two colour-singlet pairs -- were
degenerate in energy e.g. as in a square geometry. The main reasons for
studying other geometries were: 1) to investigate further this point and
2) to give a representative set of geometries illustrating the different
possibilities for arranging four quarks. The latter would then serve
as a very stringent test of models attempting to describe four
interacting quarks -- provided such models involve approximations analogous
to those contained in the corresponding lattice calculation e.g. static
quarks and SU(2).

Interest in the tetrahedral geometry arises, since in this case all three
partitions are degenerate in energy. For completeness, a series of related
geometries, where the four quarks have the coordinates
$(0,0,0),(r,0,d),(0,d,d)$ and $(r,d,0)$, are now studied. Preliminary results
are given in table~1.

\begin{table}[hbt]
\caption{The ground and first excited state energies $(E_{0,1})$. The results
for $r=0$ are from ref.[4] for the $1\times 1$-square using 2 partitions
and those for $r=d$ for the Tetrahedron.
All energies are given in terms of
lattice units at $\beta=2.4$, which corresponds to an inverse
lattice spacing $a^{-1}\approx 1.7$ GeV. }

\begin{tabular}{|c|c|c|c|} \hline
(d/a,r/a)&(1,0)&(1,1)&(1,2) \\ \hline
$E_0$ &--0.0696(2)&--0.016(2)&--0.003(1)\\
$E_1$ &--&--0.016(2)&0.265(2) \\ \hline
(d/a,r/a)&(2,1)&(2,2)&(2,3)  \\ \hline
$E_0$&--0.043(3)&--0.020(1)&--0.008(2) \\
$E_1$&0.085(2)&--0.020(1)&0.147(2) \\ \hline
(d/a,r/a)&(3,2)&(3,3)&(3,4)\\ \hline
$E_0$ &--0.041(1)&--0.028(1)&--0.010(2)\\
$E_1$ &0.048(2)&--0.028(1)&0.105(5) \\ \hline
(d/a,r/a)&(4,3)&(4,4)&(4,5) \\ \hline
$E_0$&--0.039(2)&--0.036(1)&--0.012(2) \\
$E_1$&0.03(1) &--0.036(1)&0.089(4) \\ \hline
\end{tabular}
\end{table}

Here  two points are noteworthy:

1) For the tetrahedral case (i.e. $r=d$), the ground state energy $(E_0)$
 and the first excited state energy ($E_1$) are \underline{degenerate}.
As will be seen in section 3) below, this degeneracy could have a natural
explanation using a simple model.

2) As $r$ becomes small, the results match on to the earlier ones in
ref.\cite{GMS} for $r=0$ -- the $d\times d$ square geometry, where
only the two degenerate quark partitions were employed to give the binding
energies. For example, with $(d,r)$= (2,1) and (3,2) the values of $E_0$
are  the same when using either the two degenerate
partitions or the complete set of three. However, the
results for $E_1$ with two partitions  get less accurate
as $r$ increases giving 0.086(2) and 0.053(4) for the above values of $(d,r)$.
This gives added confidence that, for squares, the earlier use of
two partitions was certainly adequate.

\section{EXTRACTION OF BINDING ENERGIES USING THE FULL CORRELATION MATRIX}

In refs.\cite{GMP2,GMS} the energies were extracted by first solving
for the eigenvalues of
\begin{equation}
\label{WABC}
W^T_{mn} a^T_n = \lambda^{(T)}_i W^{T-1}_{mn} a^T_n.
\end{equation}
Here, for euclidean time differences $T$, the $W^T_{mn}$ are the values of
Wilson loops with $m,n=1,2,3$ for the
2-quark energies (i.e. 3 fuzzing levels 12,16,20) and $m,n=1,2$ or $1,2,3$
for the 4-quark energies depending on the number of partitions used -- all
with fuzzing level 20.
Knowing $\lambda^{(T)}_i$, the 2-and 4-quark energies are given by the
plateau as $E_i =-\lim_{T \rightarrow \infty}\ln\left( \lambda^{(T)}_i\right)$.
In practice, since it is the 4-quark \underline{binding} energy that is of
prime interest, the plateau is extracted directly for the difference of the
4- and 2-quark energies. These differences are quoted in table 1.
 The errors are then estimated by bootstrapping this procedure. As discussed in
ref.\cite{MM}, this technique does not make full use of the available
Monte Carlo data, since it disregards correlations between different values
of $T$. A more complete fit minimizes the following function
\begin{equation}
\sum_{T,T',P,P'}D(T'P')C^{-1} D(TP),
\end{equation}
where $D(T P)=\left[F(T P)-W(T P)\right].$
Here $W$ is the average of the Wilson loop data and $F$ is a function to
be fitted to this data. Both of these are dependent on $T$ and also $P$ -- an
index denoting the fuzzings or partitions involved i.e. correlations among
the data in both $T$ and in $P$ are now included.
$C(T'T,P'P)$ is the full correlation matrix among the measured Wilson
loop data samples.
Since inverting a large matrix with limited data samples is unstable,
we need to  approximate $C$ by some model e.g.

$C(T'T,P'P)=A(P'P)\exp(-a|T'-T|)$

\noindent would be one particularly simple
choice. Usually $F$ has a form such as $\sum_i a_i(P) \exp(-m_i T)$, where the
$m_i$ are then interpreted as the energies ($E_i$) of the system.
Such an analysis is now being carried out. Preliminary results in table 2
indicate that the earlier energies and their error estimates
were indeed correct.

\begin{table}[hbt]
\caption{ Comparison between the energies $E_i(C)$ extracted using the full
correlation matrix  and the earlier results $E_i(P)$ with the plateau method.
These results are for quarks on the corners of rectangles with sides (d,r).}

\begin{tabular}{|c|c|c|c|} \hline
(d,r)&(2,2)&(2,3)&(2,4) \\ \hline
$E_0(C)$ &--0.0590(3)&--0.005(1)&0.0006(7)\\
$E_0(P)$ &--0.0585(4)&--0.006(1)&--0.0004(4) \\ \hline
$E_1(C)$ &0.141(1)&0.325(2)&0.514(1)\\
$E_1(P)$ &0.143(1)&0.324(1)&0.512(1)\\ \hline
\end{tabular}
\end{table}

\section{A MODEL FOR UNDERSTANDING 4-QUARK ENERGIES}

In refs.\cite{GMP}-\cite{GMS} a model for attempting to understand the
above lattice  results was introduced. This amounts to comparing the lattice
energies with  the eigenvalues [$E_i(f)$] from
\begin{equation}
\label{Hamf}
\left[{\bf V}(f)-E_i(f) {\bf N}(f)\right]\Psi_i=0,
\end{equation}
with

\[{\bf N}(f)=\left(\begin{array}{ll}
1&f/2\\
f/2&1\end{array}\right),\]
\begin{equation}
{\bf V}(f)=\left(\begin{array}{cc}
v_{13}+v_{24} & fV_{AB}\\
fV_{BA}&v_{14}+v_{23}\end{array}\right).
\end{equation}
Here $V_{AB}=V_{BA}=$
\begin{equation}
\frac{1}{2}\left(v_{13} +v_{24} +v_{14}+v_{23} - v_{12}-v_{34} \right)
\end{equation}
as expected from a ${\bf \tau}_i.{\bf \tau}_j$ colour-vector potential.
For squares of a given size $d\times d$, one value
of $f(d)$ gives a reasonable fit to both $E_0(d)$ and $E_1(d)$.
However, for the tetrahedral geometry, since all the $v_{ij}$ are equal,
the two eigenvalues of eq.(\ref{Hamf}) are degenerate at \underline{zero}
binding energy. To push this
degeneracy to some other energy could be explained by including a
colour-scalar interquark potential in addition to the colour-vector
potential used to derive eq.(5).

Further support for the above model comes from ref.\cite{Lang}, where it is
shown explicitly that perturbation theory upto fourth order in the quark-gluon
coupling does indeed correspond to the $f=1$ limit.

Of course, the above replacement of the $E_{0,1}$ by $f$ is not particularly
useful unless $f$ -- a function of all the four quark coordinates -- can
be parametrized in some convenient and simple manner.
So far, the most successful appear to be the forms
\begin{equation}
f=f_c\exp[-\alpha b_0 A-\gamma \sqrt{b_0} P],
\end{equation}
where $b_0$ is the string tension dictated by the 2-quark
potential,  $A$ is the minimal area of the surface
bounded by the straight lines connecting the quarks, $P$ is a perimeter
connecting the four quarks  and $\alpha, \gamma$ are
phenomelogical constants adjusted to fit the lattice data. In this way, the
original
rapid variations of $E_{0,1}$ with geometry can be first described in terms of
a smoothly varying function $f$, which can then be expressed in terms of the
constants $\alpha, \gamma$. In practice, two specific parametrizations have
been attempted. The first \cite{FGM} assumes $f_c=1$ -- consistent with the
expectation that $f\rightarrow 1$ as $A$ and $P$ become small i.e. in the
perturbative limit. In addition, guided by arguments based on the lattice
cubic symmetry,
$P$ is dependent on the underlying
lattice -- not just the quark positions -- and so, being a lattice artefact,
must be dropped in the continuum limit. In the second parametrization, $P$ is
defined simply in terms of the quark coordinates by the straight lines joining
the quarks. However, $f_c$ is taken to be $\beta$ dependent. A preliminary
fit gives $f_c(\beta =2.4)=0.88(2)$ and $f_c(\beta =2.5)=0.94(2)$ i.e.
apparently $f_c\rightarrow 1$ as $\beta$ increases -- again as expected in
the perturbative limit.
The continuum form of $f$ then is taken to have $f_c=1$, with $A$ and $P$
defined purely by the four quark positions. At present, it is not clear which
of these two parametrizations is to be preferred.

The authors wish to thank Dr.J.E.Paton for many useful comments at all stages
of this work. They also  wish to acknowledge that these  calculations were
performed on the Helsinki CRAY X-MP and the RAL(UK) CRAY Y-MP.

\end{document}